\documentclass[envcountsame,runningheads,11pt]{llncs}
\usepackage{a4wide}
\usepackage{xspace}
\usepackage{amsmath}
\usepackage{amssymb}
\usepackage{marvosym}
\usepackage{graphicx}
\newcommand{\reduceq}{\preceq}

\newcommand{\IR}{\mathbb{R}}
\newcommand{\IRc}{\mathbb{R}_c}

\newcommand{\IQ}{\mathbb{Q}}

\newcommand{\IN}{\mathbb{N}}
\newcommand{\dom}{\operatorname{dom}}

\newcommand{\cf}[1]{\mathbf{1}_{#1}}

\newcommand{\myjoin}{\wedge}  
\newcommand{\myjump}[1]{\widehat{#1}}
\DeclareMathOperator*{\mysup}{\smash{\sup}}
\DeclareMathOperator*{\myinf}{\smash{\inf}}

\newcommand{\calO}{\mathcal{O}}
\newcommand{\calM}{\mathcal{M}}
\newcommand{\calNP}{\mathcal{NP}}

\newcommand{\person}[1]{\textsc{#1}}
\newcommand{\aname}[1]{\textsf{#1}}
\newcommand{\mycite}[2]{{\rm\cite[\textsc{#1}]{#2}}}
\newcommand{\myto}{\!\to\!}

\newcommand{\ball}{B}
\newcommand{\closure}[1]{\overline{#1}}
\newcommand{\interior}[1]{\overset{\circ}{#1}}
\newcommand{\cball}{\closure{\ball}}
\newcommand{\myrho}{\rho}
\newcommand{\myl}{{\scriptscriptstyle<}}
\newcommand{\myg}{{\scriptscriptstyle>}}
\newcommand{\myrhol}{\myrho_{\raisebox{0.2ex}{$\myl$}}}
\newcommand{\myrhog}{\myrho_{\raisebox{0.2ex}{$\myg$}}}
\newcommand{\myrhon}{\myrho_{\text{Cn}}}

\newcommand{\myrhob}{\myrho_{\text{b},2}}
\newcommand{\myrhoh}{\myrho_{\text{H}}}

\newcommand{\psiL}[1]{\psi^{\hspace*{-0.7pt}#1}_{\!\raisebox{0.2ex}{$\myl$}}}

\newcommand{\thetal}{\theta_{\raisebox{0.1ex}{$\scriptscriptstyle<$}}}
\newcommand{\cthetaG}[1]{{\vphantom{\theta}\smash{\overline{\theta}}}^{\hspace*{-0.5pt}#1}_{\raisebox{0.5ex}{$\myg$}}}

\newcommand{\psig}{\psi_{\!\raisebox{0.2ex}{$\myg$}}}
\newcommand{\Open}{\calO}
\newcommand{\Oracle}{\calO}
\newcommand{\Alphabet}{\{0,1\}}
\newcommand{\AlphabetCup}[1]{\{0,1,#1\}}
\newcommand{\KleeneS}{\Sigma}

\newcommand{\KleeneD}{\Delta}
\newcommand{\BorelS}{\mathbf{\Sigma}}
\newcommand{\BorelP}{\mathbf{\Pi}}

\newcommand{\VascoSet}[1]{\delta_{\BorelS_{#1}(X)}}
\newcommand{\VascoFun}[1]{\delta_{\BorelS_{#1}(X\to\IR)}}

\newcommand{\myASCII}[1]{\text{\raisebox{0.4ex}{\fbox{\rule{0pt}{0.6ex}\smash{\raisebox{-0.4ex}{\tt\hspace*{-0.4ex}#1\hspace*{-0.5ex}}}}}}\,\xspace}
\newcommand{\myCR}{\myASCII{CR}}\newcommand{\myBS}{\myASCII{BS}}

\newcommand{\myedit}{\raisebox{-0.2ex}{\includegraphics[width=1.55ex]{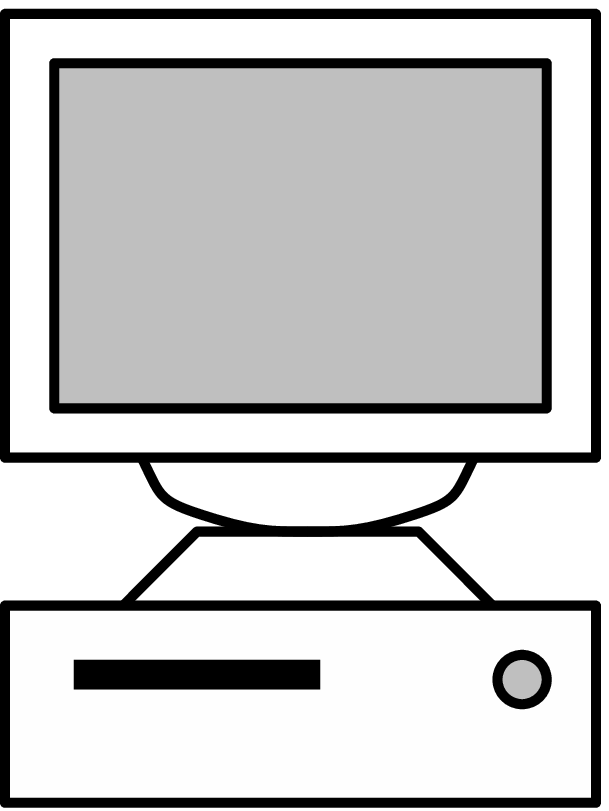}}}
\newcommand{\COMMENTED}[1]{}

\spnewtheorem{observation}[theorem]{Observation}{\bfseries}{\itshape}
\spnewtheorem{fact}[theorem]{Fact}{\bfseries}{\itshape}
\spnewtheorem{myclaim}[theorem]{Claim}{\bfseries}{\itshape}
\spnewtheorem{scholium}[theorem]{Scholium}{\bfseries}{\itshape}

\begin{document}
\setcounter{secnumdepth}{3}
\title{Revising Type-2 Computation \ and \ Degrees of Discontinuity}
\titlerunning{Revising Type-2 Computation \ and \ Degrees of Discontinuity}
\author{Martin Ziegler\thanks{Supported by \textsf{JSPS} grant \texttt{PE\,05501}.
The author wishes to express further gratitude to his Japanese
host professor \person{Hajime ISHIHARA} for exuberant 
assistance and latitude!\protect\label{f:Thanks}}}
\authorrunning{Martin Ziegler, JAIST}
\institute{Japan Advanced Institute of Science and Technology, \\
School of Information Science, 1-1 Asahidai, Nomi, Ishikawa 923-1292}
\maketitle
\def\thefootnote{\fnsymbol{footnote}}
\addtocounter{footnote}{3}
\begin{abstract}
By the sometimes so-called \emph{Main Theorem} of Recursive Analysis,
every computable real function is necessarily continuous.
Weihrauch and Zheng (TCS\,2000), Brattka (MLQ\,2005), and Ziegler 
(ToCS\,2006)
have considered different relaxed notions of computability
to cover also discontinuous functions. The present work compares
and unifies these approaches. This is based on the concept of
the \emph{jump} of a representation: both a TTE--counterpart 
to the well known recursion-theoretic jump on 
Kleene's Arithmetical Hierarchy of hypercomputation:
and a formalization of revising computation in the
sense of Shoenfield.

We also consider Markov and Banach/Mazur oracle--computation of discontinuous functions
and characterize the computational power of Type-2 nondeterminism
to coincide with the first level of the Analytical Hierarchy.
\end{abstract}
\section{Introduction} \label{s:Intro}
Every computable real function $f$ is necessarily continuous!

Computability here refers to effective $(\myrho\myto\myrho)$--evaluation 
in the sense of $x$ input to a Turing machine by means of a $\myrho$--name,
that is a fast converging sequence of rationals $(q_n)$; and $y=f(x)$ output 
in form of a similar sequence $(p_m)$. Equivalently:
the pre-image $f^{-1}[V]$ of an open set $V\subseteq\IR$ is open ;
and the mapping $V\mapsto f^{-1}[V]$ is effective in the sense that,
giving an enumeration of (the centers and radii of) 
open rational balls exhausting $V$, a
Turing machine can output a similar list exhausting $f^{-1}[V]$.
This amounts to $(\thetal\myto\thetal)$--computability of $V\mapsto f^{-1}[V]$.
\\
How can we relax this notion to include also discontinuous functions
$f:X\to\IR$? \footnote{Here and in the sequel, $X$ denotes a fixed recursively open 
(i.e. $\thetal$--computable) subset of $\IR^D$ for some $D\in\IN$.}
\begin{enumerate}
\item[i)]
  A representation (and thus a computability notion)
  for $L^2$--functions or distributions is
  devised easily and naturally \cite{Ning}; but 
  evaluation $x\mapsto f(x)$ thereon is 
  neither effective nor mathematically well-defined.
\item[ii)]
  Granting a Type-2 machine access to an oracle like, say, 
  the Halting problem increases its recursion-theoretic power 
  but does not lift the topological restriction
  to continuous real functions; see e.g. \mycite{Lemma~8}{TOCS}.
\item[iii)]
  \person{Weihrauch} and \person{Zheng} (2000) have considered
  $(\myrho\myto\myrhol)$--computable functions where the output
  representation $\myrhol$ encodes $y=f(x)\in\IR$ as a rational sequence
  $(p_m)$ with $y=\sup_m p_m$. Such functions are in general only
  lower \emph{semi-}continuous  \cite{SemiTCS}, that is, the pre-image
  $f^{-1}[V]$ is open for every $V=(y,\infty)$. As a matter of fact,
  $f$ is $(\myrho\myto\myrhol)$--computable ~if and only if~
  $a\mapsto f^{-1}[(y,\infty)]$ is $(\myrhog\myto\thetal)$--computable
  \mycite{Theorem~4.5}{SemiTCS}.
\item[iv)]
  Motivated by (a different) work of \person{Zheng} and \person{Weihrauch} (2001),
  \cite{TOCS} introduced representations $\myrho'$, $\myrhol'$, $\myrho''$, \ldots,
  $\myrho^{(d)}$, $\myrhol^{(d)}$ weakening $\myrho$ and $\myrhol$. 
  A real number $x$ is 
  $\myrho$--computable \emph{relative} to the Halting problem $\emptyset'$
  ~if and only if~ it is $\myrho'$--computable \cite{Ho}.
  More generally, $x$ is $\myrho$--computable \emph{relative} to $\emptyset^{(d)}$
  ~if and only if~ $x$ is $\myrho^{(d)}$--computable \cite{Xizhong};
  similarly for $\myrhol^{(d)}$. These representations thus 
  parallel the levels $\KleeneS_d$ of \person{Kleene}'s \textsf{Arithmetical Hierarchy}.
\item[v)]
  \person{Brattka} relaxes the pre-image mapping $V\mapsto f^{-1}[V]$ 
  from being open and effectively open and
  instead considers $\BorelS_d$--measurability \cite{EffBorel}.
  This condition requires that $f^{-1}[V]$
  be a $\BorelS_d$ set in \person{Borel}'s topological hierarchy. 
  For its ground level $\BorelS_1(X)$ of open subsets of $X$, he thus recovers classical 
  continuity; $\BorelS_2(X)$ consists of the $F_\sigma$ sets,
  and so on. The mapping $V\mapsto f^{-1}[V]$ must furthermore be effective
  in the sense that, given a $\thetal$--name of $V$, a Type-2 machine must
  be able to obtain a name of $f^{-1}[V]$ in terms of the natural representation
  $\VascoSet{d}$ of $\BorelS_d(X)$; ~ $\VascoSet{1}\equiv\thetal$.
\item[vi)]
  Real nondeterminism had been introduced in \mycite{Section~5}{CIE,TOCS}.
  A corresponding machine computing $y=f(x)$ may make a binary
  choice at each step, as long as any \emph{in}finite output sequence $(q_n)$ 
  constitutes a $\myrho$--name of $y$.
  This notion has been shown to include all $(\myrho\myto\myrho^{(d)})$--computable functions
  \mycite{Theorem~28}{TOCS}.
\end{enumerate}
Notice that proceeding from $(\myrho\myto\myrho)$--computability
to $(\myrho\myto\myrho^{(d)})$--computability amounts to 
weakening the information to be output for the \emph{values}
(image) of the function $f$ under consideration;
whereas proceeding from effective $\BorelS_1$--measurability
(equivalent to $(\myrho\myto\myrho)$--computability)
to, say, effective $\BorelS_{d+1}$--measurability
amounts to weakening the encoding on the \emph{pre-}image
side (i.e. the domain) of $f$.

\subsection{Overview}
The present work unifies and extends approaches iii), iv), and v) above.
Some main results are collected in the following
\begin{theorem} \label{t:Main}
Fix a function $f:X\to\IR$ and $d\in\IN$.
\begin{enumerate}
\item[a)]
  $f$ is $(\myrho\myto\myrho^{(d-1)})$--computable
  ~if and only if~
  it is effectively $\BorelS_d$--measurable.
\item[b)]
  $f$ is $(\myrho\myto\myrhol^{(d-1)})$--computable
  ~if and only if~ the mapping 
  $\IR\ni y\mapsto f^{-1}[(y,\infty)]\in\BorelS_d(X)$
  is well-defined and $(\myrhog\myto\VascoSet{d})$--computable.
\item[c)]
  There exists a nondeterministically computable total real function
  which is not $(\myrho\myto\myrho^{(d)})$--computable for any $d\in\IN$ whatsoever.
\end{enumerate}
\end{theorem}
In particular, weakly evaluable functions (in the sense of iv) range
arbitrarily high on \person{Borel}'s taxonomy of discontinuity
but are strictly succeeded by nondeterminism (vi).
Theorem~\ref{t:Main}a) also gives one explanation
for the dominance in \cite{EffBorel}
of the Borel classes $\BorelS_d$ over the (seemingly more
symmetric ones) $\Delta_d$.

Claims~a) and b) in the above theorem turn out to actually hold
even uniformly in $f$. To this end, we introduce 
in Section~\ref{s:Main}
the notion of $\BorelS_d$--\emph{semi}measurability 
and a representation for according functions: 
a generalization unifying both \cite{Xizhong} and \cite{EffBorel}.
The central concept in the present work is that of the 
\emph{jump} $\alpha'$ of a representation $\alpha$
(Section~\ref{s:Jump}). 
For the case $\alpha=\myrho$, it coincides with the notion
from \cite{TOCS} and simplifies the proofs therein.

Motivated by \emph{revising} computation,
Section~\ref{s:Revising} considers an equally natural but different
kind of jump operator on representations. The power of Type-2 Nondeterminism
\mycite{Section~5}{CIE,TOCS} is the topic of Section~\ref{s:Nondette}.
And before concluding, we also briefly dive into oracle--supported
Markov and Banach/Mazur computability (Section~\ref{s:Markov}).

\section{The \emph{Jump} of a Representation} \label{s:Jump}
\person{Ho} has shown that a real number $x$ is $\myrho$--computable
(that is admits effective approximations by a fast converging rational sequence)
\emph{relative} to the Halting problem $\emptyset'$ 
~if and only if~ $x$ is the (unconditional) limit of a computable
rational sequence \mycite{Theorem~9}{Ho}. This has suggested the
alternative name $\myrho'$ for the \emph{naive} Cauchy representation
encoding $x$ as an ultimately converging rational sequence.
Another example, \person{Brattka} has weakened (and extended)
the representation $\thetal\equiv\VascoSet{1}$ for open sets to $\VascoSet{d}$ mentioned above.
The present section unifies these and several other notions.

We start with Cantor space $\Alphabet^\omega$ which is usually and
canonically represented by the identity $\imath$ 
\mycite{Definition~3.1.2.1}{Weihrauch}.

\begin{definition} \label{d:CantorJump}
Let the representation $\imath':\subseteq\Alphabet^\omega\to\Alphabet^\omega$
encode an infinite string $\bar\sigma\in\Alphabet^\omega$
as (the pairing of) 
a sequence of infinite strings ultimately converging to $\bar\sigma$.
\end{definition}
This amounts to the \emph{naive} Cauchy representation of the
effective metric Cantor space \mycite{Section~6}{Naive}.
An $\imath'$--name for $(\sigma_n)_{_n}$ is thus (an $\imath$--name for)
some $\big((\tau_{\scriptscriptstyle\langle n,m\rangle})_{_{\scriptstyle n}}\big)_m\in\Alphabet^\omega$
such that, for each $n\in\IN$, 
$\sigma_n=\lim_{m\to\infty} \tau_{\langle n,m\rangle}$.
The name $\imath'$, 
reminiscent of the recursion-theoretic \emph{jump}, is justified 
because \textsf{Shoenfield's Limit Lemma} immediately yields

\begin{observation} \label{o:Shoenfield}
Let $\Oracle$ denote an arbitrary oracle.
An infinite string is ($\imath$--) computable
\emph{relative} to $\Oracle'$
~if and only if~ it is $\imath'$--computable
relative to $\Oracle$.
\end{observation}
Moreover we have
\begin{lemma} \label{l:CantorJump}
\begin{enumerate}
\item[a)] Every ($(\imath\myto\imath)$--) computable
  string function $F:\subseteq\Alphabet^\omega\to\Alphabet^\omega$
  is also ($\imath'\myto\imath')$--computable;
\item[b)] more precisely the \textsf{apply operator}
  $(F,\bar\sigma)\mapsto F(\bar\sigma)$ is
  $(\eta^{\omega\omega}\times\imath'\myto\imath')$--computable.
\item[c)] Every $(\imath'\myto\imath')$--continuous
  string function $F:\subseteq\Alphabet^\omega\to\Alphabet^\omega$
  is (Cantor--)continuous.
\item[d)] Whenever $\alpha:\subseteq\Alphabet^\omega\to A$ is a representation for $A$,
then so is $\alpha\circ\imath'$.
\item[e)] $\alpha\reduceq\beta$ ~implies~ $\alpha\circ\imath'\reduceq\beta\circ\imath'$.
\end{enumerate}
\end{lemma}
In b), $\eta^{\omega\omega}$ denotes a natural representation
for continuous string functions \mycite{Section~2.3}{Weihrauch}.
\begin{proof}\begin{enumerate}
\item[a)] follows from b).
\item[b)] Let $\tilde\tau_m:=F(\bar\tau_m)$ where 
  $F:\subseteq\Alphabet^\omega\to\Alphabet^\omega$ is continuous.
  Then $\lim\nolimits_{m}\tilde\tau_m = F\big(\lim\nolimits_{m}\bar\tau_m\big)$.
\item[c)] See \mycite{Section~6}{Naive}.
\item[d)] immediate.
\item[e)] Let $F$ denote a computable string function
 converting $\alpha$--names to $\beta$--names. By a), $F$
 has a computable $(\imath'\myto\imath')$--realization 
 $G:\subseteq\Alphabet^\omega\to\Alphabet^\omega$.
 This $G$ converts $(\alpha\circ\imath')$--names to
  $(\beta\circ\imath')$--names.
\qed\end{enumerate}\end{proof}
The rest of this section relates several known representations
to ones of the form $\alpha\circ\imath'$ for some $\alpha$.%

\subsection{Weak Real Representations} \label{s:JumpRho}
Recall from \mycite{Section~2}{CIE,TOCS} the following 

\begin{definition} \label{d:HyperReal} 
Consider the representations of $\IR$ where 
a real $y$ is encoded as%
\\[0.7ex]\noindent\begin{tabular}{ll}
$\pmb{\myrho}$&\textbf{:} a rational sequence $(p_m)$ 
  such that $|y-p_m|\leq2^{-m}$\hfill(i.e. fast convergence)\\[0.7ex]
$\pmb{\myrhol}$&\textbf{:} a rational sequence $(p_m)$ 
  such that $y=\mysup\limits_m p_m$\hfill(i.e. lower approximation)\\[0.7ex]
$\pmb{\myrho'}$&\textbf{:} a rational sequence $(p_m)$ 
  such that $y=\lim\limits_m p_m$\hfill(i.e. ultimate convergence)\\[0.7ex]
$\pmb{\myrhol'}$&\textbf{:} a rational sequence $(p_m)$ 
  such that $y=\mysup\limits_m \myinf\limits_n p_{\langle m,n\rangle}$\qquad
(equivalently: $\liminf$)\\[0.7ex]
$\pmb{\myrho''}$&\textbf{:} a rational sequence $(p_m)$ 
  such that $y=\lim\limits_m \lim\limits_n p_{\langle m,n\rangle}$\\[0.7ex]
$\pmb{\myrhol''}$&\textbf{:} a rational sequence $(p_m)$ 
  such that $y=\mysup\limits_{m}\myinf\limits_{n}\mysup\limits_{k}
  p_{\langle m,n,k\rangle}$%
\\[-1.5ex]$\vdots$\\[-0.7ex]
$\pmb{\myrho^{(d)}}$&\textbf{:} a rational sequence $(p_m)$ 
  such that $y=\lim\limits_{n_1} \lim\limits_{n_2}\lim\limits_{n_3}\cdots\lim\limits_{n_d} 
  p_{\langle n_1,n_2,\ldots,n_d\rangle}$%
\end{tabular}
\end{definition}
($\myrho'$ of course coincides with the well-known
\emph{naive} Cauchy-representation $\myrhon$.)
These encodings constitute a hierarchy 
\[ \myrho \;\;\reduceq\;\;
 \myrhol\;\;\reduceq\;\;\myrho'\;\;\reduceq\;\;
 \myrhol'\;\;\reduceq\;\;\myrho''
 \;\;\reduceq\;\;\myrhol''\;\;\reduceq\;\;\ldots\;\;\reduceq\;\;
 \myrho^{(d)}\;\;\reduceq\;\;\myrhol^{(d)}\;\;\reduceq\;\;\ldots \]
of representations introduced in \mycite{Section~2.2}{TOCS}. 
This hierarchy correspond to---and is in particular as strict as---\person{Kleene}'s 
\textsf{Arithmetical Hierarchy} of hypercomputation
\newcommand{\mytmpspace}{\hspace*{4.8pt}}
\[ \KleeneD_1\mytmpspace\subsetneq\mytmpspace\KleeneS_1\mytmpspace\subsetneq\mytmpspace\KleeneD_2\mytmpspace\subsetneq\mytmpspace\KleeneS_2\mytmpspace\subseteq\mytmpspace
\KleeneD_3\mytmpspace\subseteq\mytmpspace\KleeneS_3\mytmpspace\subseteq\mytmpspace\ldots\mytmpspace\subseteq\mytmpspace\KleeneD_{d+1}\mytmpspace\subseteq\mytmpspace\KleeneS_{d+1}\mytmpspace\subseteq\mytmpspace\ldots \]
in the following way:
A real number $y$ is $\myrho^{(d)}$--computable if and only if
$y$ is $\myrho^{(k)}$--computable \emph{relative} to $\emptyset^{(d-k)}$
for some (or, equivalently, for every) $0\leq k\leq d$; and $y$ is 
$\myrhol^{(d)}$--computable if and only if
$y$ is $\myrhol^{(k)}$--computable relative to to $\emptyset^{(d-k)}$,
see \mycite{Section~7}{Xizhong}. Notice how this extends
\person{Shoenfield}'s Limit Lemma from discrete to the
continuous realm \mycite{Section~4}{Xizhong}.

\subsection{Jump of the Cauchy Representation} 
\begin{proposition} \label{p:JumpRho}
~ $\displaystyle \myrho\circ\imath'\;\equiv\; \myrho'$.
\end{proposition}
In combination with Observation~\ref{o:Shoenfield}, 
this implies \mycite{Theorem~9}{Ho}; and together with
Lemma~\ref{l:CantorJump}b) it includes \mycite{Scholium~17}{TOCS}.

\begin{proof}
A $(\myrho\circ\imath')$--name for $x\in\IR$ is (basically)
a sequence of rational sequences
eventually stabilizing (elementwise) to a fast converging Cauchy sequence
$(q_{(n,\infty)})_{_n}$;
that is a double sequence $(q_{(n,m)})$ in $\IQ$ such that 
\[ \forall n \; \exists m_0 \; \forall m\geq m_0: \quad q_{(n,m)}=q_{(n,m_0)} 
\;\wedge\; |x-q_{(n,m_0)}|\leq2^{-n} \enspace . \]
\begin{description}
\item[$\pmb{\reduceq}$:]
For each $m$, let $(q_{(1,m)},q_{(2,m)},\ldots,q_{(N_m,m)})$ denote
the longest initial part of $(q_{(1,m)},\ldots,q_{(m,m)})$ satisfying
\begin{equation} \label{e:JumpRho}
|q_{(n,m)}-q_{(n',m)}| \;\leq\; 2^{1-n}\quad\forall 1\leq n\leq n'\leq N_m \enspace .
\end{equation}
Since $(q_{(n,\infty)})_{_n}$ is a $\myrho$--name and due to the
eventual stabilization, $N_m\to\infty$ as $m\to\infty$.
Also, the sequence $(N_m)_{_m}$ is computable from the above input.
Consider the following algorithm, starting with empty output tape:

For each $m=1,2,\ldots$, 
test whether the initial parts of $q_{(\cdot,m)}$ and $q_{(\cdot,m-1)}$
up to $N_m$ coincide: 
$(q_{(1,m)},\ldots,q_{(N_m,m)})=(q_{(1,m-1)},\ldots,q_{(N_m,m-1)})$?
(For notational convenience, set $q_{n,0}:\equiv\infty$ and $N_0:=0$.)
If so, then obviously $N_m\geq N_{m-1}$; so append (the possibly empty sequence)
$(q_{(N_{m-1},m)},\ldots,q_{(N_m,m)})$ to the output.
Otherwise let $n_m$ be maximal with
$(q_{(1,m)},\ldots,q_{(n_m,m)})=(q_{(1,m-1)},\ldots,q_{(n_m,m-1)})$;
obviously $n_m<N_m$, so append 
$(q_{(n_m,m)},\ldots,q_{(N_m,m)})$ to the output in this case.

It remains to show that that yields a valid $\myrho'$--name for $x$.
Let $\epsilon=2^{1-n}$.
Then $|q_{(n,\infty)}-q_{(n',\infty)}|\leq\epsilon$ for all $n'\geq n$
because $q_{(n,\infty)}$ constitutes a $\myrho$--name.
Moreover due to stabilization, there exists some maximal
$m$ with $q_{(n,m)}\not=q_{(n,m-1)}$. 
During the phase no.$m$ corresponding to that last change,
the above algorithm will detect $n_m<N_m$ and thus output 
(a finite sequence beginning with) $q_{(n,m)}$.
Moreover as $q_{(n,\cdot)}$ afterwards does not change anymore,
all elements $q_{(n',m')}$ appended subsequently
will have $n'\geq n$ and $m'\geq m$; 
in fact $N_{m'}\geq n'\geq n_{m'}\geq n$,
hence $|q_{(n',m')}-q_{(n,m)}|\leq\epsilon$ 
because $q_{(n,m')}=q_{(n,m)}$ and due to Equation~(\ref{e:JumpRho}).
Therefore the output constitutes a (naive) Cauchy sequence
converging to $x$.
\item[$\pmb{\succeq}$:]
Let $(q_n)_{_n}$ be a sequence in $\IQ$ ultimately converging to $x$.
There exists an increasing sequence $(n_m)_{_m}$ in $\IN$ such that
\begin{equation} \label{e:JumpRho2}
\forall k\geq n_m: \quad |q_{n_m}-q_k|\;\leq\;2^{-m-1} \enspace .
\end{equation}
The subsequence $(q_{n_m})_{_m}$ constitutes a $\myrho$--name for $x$.
For each single $m$, Condition~(\ref{e:JumpRho2}) 
can be falsified (formally: is co-r.e. in the input).
A Turing machine is therefore able to iteratively try
for $n_m$ all integer values from $n_{m-1}$ on 
and fail only finitely often for each $m$.

Trial no.$\ell$ thus yields a sequence $(n'_{(\ell,m)})_{m\leq\ell}$
of length $\ell$
such that, for each $m$, $n'_{(\cdot,m)}$ eventually stabilizes to $n_m$
satisfying (\ref{e:JumpRho2}). By artificially extending each finite sequence
to an infinite one, we obtain a $\myrho\circ\imath'$--name for $x$.
\qed\end{description}\end{proof}

\subsection{Jump of Lower Real Representation}
Our next result includes, in view of Lemma~\ref{l:CantorJump}a+c),
\cite[\textsc{Theorem~11}a+b) and \textsc{Theorem~15}b+c)]{TOCS}
because $(\myrho\myto\myrhol)$--continuity implies lower-semicontinuity
and $(\myrhol\myto\myrhol)$--continuity requires monotonicity
\cite{SemiTCS}.

\begin{proposition} \label{p:JumpRhol}
~ $\displaystyle \myrhol\circ\imath'\;\equiv\; \myrhol'$.
\end{proposition}
\begin{proof}
A $(\myrhol\circ\imath')$--name for $x\in\IR$ amounts to
a sequence of rational sequences
eventually stabilizing (elementwise) to a sequence
approaching $x$ from below,
that is a double sequence $(q_{(n,m)})$ in $\IQ$ such that 
\[ \forall n \; \exists m_0=m_0(n) \; \forall m\geq m_0: \quad q_{(n,m)}=q_{(n,m_0)} 
\;\wedge\; x=\mysup_n q_{(n,m_0(n))} \enspace . \]
\begin{description}
\item[$\pmb{\reduceq}$:] Since the limit (which exists) 
coincides with the least accumulation point, we have%
\[ x\quad=\quad \mysup_n \lim_m q_{(n,m)} \quad=\quad
  \mysup_n \mysup_j \myinf_{m\geq j} q_{(n,m)}
\quad=\quad \mysup_{\langle n,j\rangle} \myinf_m \left\{ \begin{array}{ll}
 q_{(n,m)} &:\; m\geq j \\ \infty &:\; m<j \end{array} \right. \]
deduced a $\myrhol'$--name for $x$.
\item[$\pmb{\succeq}$:]
Let $(q_{(n,m})$ be the given double
sequence in $\IQ$ with $x=\mysup_n\myinf_m q_{(n,m)}$.
We may suppose that all single sequences $q_{(n,\cdot)}$, $n\in\IN$, 
are monotonically nonincreasing; and that the single sequence 
$\big(\myinf_m q_{(n,m)}\big)_{_n}$ is nondecreasing:
by proceeding (in either order!) from $q_{(n,m)}$ to 
$\min_{k\leq m} q_{(n,k)}$
and to $\max_{\ell\leq n} q_{(\ell,m)}$, respectively.
Moreover one can assert each single sequence $q_{(n,\cdot)}$ to 
eventually stabilize, thus yielding a $\myrhol\circ\imath'$--name of $x$:
Consider for $m\in\IN$ the function $\lfloor\,\cdot\,\rfloor_m:\IQ\to\IQ$ 
mapping every rational to the next lower
dyadic rational having denominator $2^{-m}$;
formally: $a/b\mapsto\frac{\lfloor a\cdot 2^m/b\rfloor}{2^m}$
where $\lfloor\,\cdot\,\rfloor=\lfloor\,\cdot\,\rfloor_0$ 
denotes the usual floor function on integers.
Then proceeding from $q_{(n,m)}$ to $\lfloor q_{(n,m)}\rfloor_{_m}$
satisfies this requirement without affecting $x=\mysup_n\myinf_m q_{(n,m)}$.
\qed\end{description}\end{proof}

\subsection{Jump of the Weierstra\ss{} Representation}
The limit of a uniformly converging sequence of polynomials 
is of course continuous again. \person{Weierstra\ss} has shown that
the converse holds as well: Any continuous function on a compact set
is the uniform limit of a sequence of polynomials.
This leads to the Weierstra\ss{} Representation 
$[\myrho\myto\myrho]$ of the class $C(K)$
of continuous functions $f:K\to\IR$ for compact $K:=[0,1]^D$:
a name of $f\in C(K)$ is (an encoding of the degrees and coefficients of)
a sequence of polynomials $P_n\in\IQ[X]$ with
\begin{equation} \label{e:Weierstrass}
\sup_{x\in K} |f(x)-P_n(x)|
\;=:\;\|f-P_n\| \;\overset{!}{\leq}\; 2^{-n} \enspace .
\end{equation}
By the famous \textsf{Effective Weierstra\ss{} Theorem},
it is equivalent to several other natural representations
of $C(K)$ \mycite{Section~6.1}{Weihrauch}.
\cite[\textsc{Lemma~12}b]{CIE} and \mycite{Lemma~22}{TOCS}
employ a representation $[\myrho\myto\myrho]'$ for
$C(K)$ where the required \emph{fast} uniform convergence 
bound $2^{-n}$ in Equation~(\ref{e:Weierstrass}) is weakened
to `ultimate' uniform convergence $\|f-P_n\|\to0$.
This kind of \emph{naive} Weierstra\ss representation,
too, results from a jump:

\begin{proposition} \label{p:Weierstrass}
~ $\displaystyle [\myrho\myto\myrho]\circ\imath'\;\equiv\; [\myrho\myto\myrho]'$.
\end{proposition}
This result includes \mycite{Theorem~16}{Ho}.
The proof proceeds similarly to that of Proposition~\ref{p:JumpRho}
because Equations~(\ref{e:JumpRho}) and (\ref{e:JumpRho2})
are still decidable and co-r.e. when replacing rational numbers
$q$ with rational polynomials $Q$ and absolute value $|q|$
with maximum norm $\|Q\|$:
\begin{fact}
Given $q_0,\ldots,q_m,b\in\IQ$ (in binary encoding, say),
~$\sup_{0\leq x\leq 1} |q_0+q_1x+\ldots+q_mx^m|=b$~
is decidable:
by virtue of constructive root bounds, 
see e.g. \cite{Mignotte}.
\end{fact}

\subsection{Iterated Jumps} 
Climbing up in \person{Kleene}'s Arithmetical Hierarchy
corresponds to iterated jumps of the Halting problem.
We proceed similarly with our hierarchy of representations:

\begin{definition} \label{d:Iterated}
Let $\imath^{(d+1)}:=\imath^{(d)}\circ\imath'=\imath'\circ\imath^{(d)}$.
\end{definition}
Straight forward
inductive application of Observation~\ref{o:Shoenfield} shows
that $\imath^{(d)}$--computability is equivalent to
$\imath$--computability relative to $\emptyset^{(d)}$.
If $F$ and $G$ are partial $(\imath\myto\imath')$--computable string functions,
then their composition $G\circ F$ is $(\imath\myto\imath'')$--computable
by Lemma~\ref{l:CantorJump}a). 

\begin{theorem} \label{t:JumpRhoD}
For each $d\in\IN$, it holds
$\myrho^{(d)}\equiv\myrho\circ\imath^{(d)}$
and $\myrhol^{(d)}\equiv\myrhol\circ\imath^{(d)}$.
\end{theorem}
\begin{proof}
The induction start $d=1$ has been treated in Propositions~\ref{p:JumpRho}
and \ref{p:JumpRhol}, respectively.
Since a $\myrho^{(d+1)}$--name of $x\in\IR$ is the join of 
$\myrho^{(d)}$--names of elements $x_n$ with $x=\lim_n x_n$,
Proposition~\ref{p:JumpRho} 
together with Lemma~\ref{l:CantorJump}e)
also provides the induction step;
similarly for $\myrhol^{(d+1)}$.
\qed\end{proof}
As a consequence, we obtain the following extensions of
\cite[\textsc{Theorems~11} and \textsc{15}]{TOCS}:
\begin{corollary} \label{c:MainTheorems}
Fix $f:X\to\IR$.
\begin{enumerate}
\item[a)]
  If $f$ is $(\myrho^{(d)}\myto\myrho^{(d)})$--continuous,
  then it is continuous.
  If $f$ is $(\myrho^{(d)}\myto\myrho^{(d)})$--computable,
  then it is also $(\myrho^{(d+1)}\myto\myrho^{(d+1)})$--computable,
\item[b)]
  If $f$ is $(\myrho^{(d)}\myto\myrhol^{(d)})$--continuous,
  then it is lower semi-continuous.
  If $f$ is $(\myrho^{(d)}\myto\myrhol^{(d)})$--computable,
  then it is also $(\myrho^{(d+1)}\myto\myrhol^{(d+1)})$--computable.
\item[c)]
  If $f$ is $(\myrhol^{(d)}\myto\myrhol^{(d)})$--continuous,
  then it is monotonically nondecreasing.
  If $f$ is $(\myrhol^{(d)}\myto\myrhol^{(d)})$--computable,
  then it is also $(\myrho^{(d+1)}\myto\myrhol^{(d+1)})$--computable.
\end{enumerate}
\end{corollary}
The proof of \mycite{Theorem~11}{TOCS} covers as many as 
five pages of text and treated only very small values of $d$.
Now it boils down to a mere application of
Lemma~\ref{l:CantorJump}a+c) inductively in $d$.

\subsection{Borel Set Representations} \label{s:JumpThetal}
The representation $\thetal$ encodes an open subset $U$ of $X$
as a list of (centers and radii) of open rational balls
exhausting $U$.
For a topological space $X$, the \textsf{Borel Hierarchy} starts
with the class $\BorelS_1(X)$ of open subsets $U$ of $X$ and
proceeds inductively from $\BorelS_d(X)$ to the class
$\BorelS_{d+1}(X)$ of countable unions $\bigcup_m (X\setminus S_m)$
over complements of sets $S_m$ from $\BorelS_d(X)$.
\person{Brattka} has renamed $\thetal$ to $\VascoSet{1}$
and generalized it to higher order Borel sets:

\begin{definition} \label{d:HyperBorel}
Consider the following representations of Borel subsets of $X$:
\\\noindent\begin{tabular}{lll}
$\pmb{\VascoSet{1}}$ & encodes $U\in\BorelS_1(X)$ &
as a list $B_m$ of open rational balls such that $U=\bigcup\limits_m B_m$ \\[0.7ex]
$\pmb{\VascoSet{2}}$ & encodes $S\in\BorelS_2(X)$ &
as a list $B_m$ of open rational balls such that \\
&&$S \quad=\quad \bigcup\nolimits_m \big( X\setminus \bigcup\nolimits_n B_{\langle m,n\rangle}\big)$\\[-2.5ex]
\vdots\\
$\pmb{\VascoSet{d}}$ & encodes $S\in\BorelS_{d}(X)$ &
as (the join of) $\BorelS_{d-1}$--names of sets $S_m\in\BorelS_{d-1}(X)$ \\
&&such that  $S=\bigcup_m (X\setminus S_m)$.
\end{tabular}
\end{definition}
It turns out that these natural representations are related to jumps, too:

\begin{proposition} \label{p:HyperBorel}
~ $\displaystyle \thetal\circ\imath'\;\reduceq\; \VascoSet{2}\big|^{\BorelS_1(X)}$.
\end{proposition}
Recall that $\BorelS_1(X)$ denotes the class of open subsets of $X$
which $\thetal\equiv\VascoSet{1}$ is a representation for.
Of course the restriction of $\VascoSet{2}$ is thus necessary
for the equivalence to make sense.
\COMMENTED{
By induction, we immediately obtain

\begin{corollary}
For each $1\leq k\leq d$, ~ 
$\VascoSet{k}\circ\imath^{(d-k)}\;\equiv\;\VascoSet{d}\big|^{\BorelS_k(X)}$.
\end{corollary}
For $k=1$, Theorem~\ref{t:Main}c) now follows with Observation~\ref{o:Shoenfield}.
}

\begin{proof}
A $(\thetal\circ\imath')$--name for $U\in\Open(X)$ consists of two rational
double sequences $(c_{(n,m)})$ and $(r_{(n,m)})$ such that
each single sequence $c_{(n,\cdot)}$ and $r_{(n,\cdot)}$, $n\in\IN$,
eventually stabilizes to some $c_{(\cdot,\infty)}$ and $r_{(\cdot,\infty)}$
where $U=\bigcup_{n\in\IN} \ball(c_{(n,\infty)},r_{(n,\infty)})$
and $\ball(c,r)$ denotes the open ball with center $c$ and radius $r$.
\\
Both representations admit effective countable unions:
apply Lemma~\ref{l:CantorJump}a) to \mycite{Example~5.1.19.1}{Weihrauch}
and see \mycite{Proposition~3.2(5)}{EffBorel}, respectively.
It therefore suffices to show them equivalent on open rational balls,
that is we may suppose w.l.o.g. $U=\ball(c_m,r_m)$ for all $m\geq m_0$.
\COMMENTED{
\begin{description}
\item[$\pmb{\reduceq}$:] Let 
} \\ So let
\[ A_n \quad:=\quad \left\{ \begin{array}{ll}
\cball\big(c_m,r_m\cdot(1-2^{-n})\big) &:\; (c_k,r_k)=(c_{k+1},r_{k+1})\forall k\geq m \\[0.7ex]
 \emptyset &:\; \text{otherwise} \end{array}\right. \enspace ,\]
so $U=\bigcup_n A_n$. Moreover the closed set 
$A_n$ can be $\psig$--computed, uniformly in $n$
and the given sequences $(c_m)$ and $(r_m)$: start generating
$\cball(\cdots)$;
if the co-r.e. condition ``$\forall k\geq m$'' eventually turns out to
fail, the machine may still revert to a $\psig$--name for $\emptyset$
by adding further negative information to the output.
Hence we obtain a $\VascoSet{2}$--name for $U$.%
\COMMENTED{
\item[$\pmb{\succeq}$:] 
For a set $S\subseteq X$ consider an enumeration of all closed
rational balls inside of $S$. In case that $S$ is open, this 
coincides with the representation $\thetal$ \mycite{Corollary~5.1.16.4}{Weihrauch};
but it can also represent the hyperspace of all regular closed sets 
$S=\closure{S^\circ}$, compare \cite{MLQ}.

Unfortunately the proof of this direction just turned out to be flawed.
I am however optimistic to repair it until the final version.
Thanks for your understanding. 
\qed\end{description}%
}\qed
\end{proof}

\COMMENTED{
\begin{lemma} \label{l:Baire}
Let $X$ denote a \textsf{Baire Space}
and $A_n\subseteq X$ a countable family of closed subsets.
If their union $U:=\bigcup_n A_n$ is open, 
then it coincides with $\bigcup_n \closure{A_n^\circ}$.
\end{lemma}
\begin{proof}
$V:=\bigcup_n \closure{A_n^\circ}$ is a subset of $U$
because $\closure{A^\circ}\subseteq\closure{A}=A$.
For the reverse inclusion ``$U\subseteq V$''
first suppose that $V=\emptyset$, that is, all $A_n$ are nowhere dense.
Then their union has empty interior $\emptyset=U^\circ=U$, too,
by the very definition of a Baire Space.
The general case $V\supseteq\emptyset$ reduces to the above
by considering closed $\tilde A_n:=A_n\setminus V$,
$\tilde U:=\bigcup_n\tilde A_n=U\setminus V$, and 
$\tilde V:=\bigcup_n\closure{A_n^\circ}=\emptyset$.
\qed\end{proof}
}
The following extends \mycite{Example~5.1.17.2}{Weihrauch}
and \cite[\textsc{Corollary~6.6}a)]{Xizhong}:
\begin{example}
For reals $a<b$, the open interval $U=(a,b)\subseteq\IR$
is $\BorelS_d$--computable ~if and only if~
$+b$ and $-a$ are both $\myrhol^{(d-1)}$--computable.
\end{example}

\begin{conjecture}
The converse of Proposition~\ref{p:HyperBorel} holds as well:
~$\displaystyle \VascoSet{2}\big|^{\BorelS_1(X)}\;\reduceq\;\thetal\circ\imath'$.
\end{conjecture}

\begin{problem}
Recalling the weak representations of regular sets
$\overline{\rule{0pt}{1.6ex}\makebox[2ex][l]{$\smash{\overset{\hspace*{1.5pt}\circ}{\rule{0pt}{0.6ex}\smash{\psiL{}}}}$}}\hspace*{1ex}$
and $\cthetaG{}$
from \mycite{Definition~3.3}{MLQ}, characterize them in terms of $\imath'$
and some known representations!
\end{problem}

\section{Revising Computation} \label{s:Revising}
This section provides some motivation and related background 
to the jump $\alpha'$ of a representation $\alpha$
as well as for a different kind of jump $\myjump\alpha$
to be introduced in Section~\ref{s:HotzJump} below.

\medskip\noindent
An important (though somewhat hidden) point in the definition of a
Type-2 machine is that its output tape be \emph{one-way};
compare e.g. \cite[top of p.15]{Weihrauch}.
This condition allows to abort a real number computation 
as soon as the desired precision is reached, knowing
that this preliminary approximation will \emph{not be reverted}.
It also is crucial for the \textsf{Main Theorem} to hold.

In the Type-1 setting, \textsf{revising computations} have been studied well.
Here a machine writes only a finite string, but it does not terminate
and may \emph{revert} its output an arbitrary finite number of times.
The model with this semantics goes under such names as 
\aname{Limiting} \cite{Gold},
\aname{Trial-and-Error} \cite{Putnam}, 
\aname{Inductive} \cite{Burgin},
or \aname{General} \cite{Schmidhuber} Turing Machines.
It is motivated by the capabilities of early display terminals
(see Section~\ref{s:Terminal} below)
as well as by \person{Shoenfield}'s Limit Lemma. 

A sequence $(\vec\sigma_n)_{_n}$ of finite strings 
(Type-1) converges (to a finite string)
~if and only if~ the sequence $\sigma_{n,i}$ of $i$-th symbols
eventually stabilizes for each $i$.
For \emph{in}finite strings (Type-2 setting) however,
one has to carefully distinguish both conditions:
symbol-wise convergence underlies Definition~\ref{d:CantorJump}
whereas overall stabilization will be required in
Definition~\ref{d:CantorHotz}.

Both appear naturally when formalizing the output displayed
by a (not necessarily terminating) program to a terminal
as explicated in Section~\ref{s:Terminal}:
They also arise as input fed to a streaming algorithm:

\subsection{Revising Input: Streams} \label{s:Stream}
Many practical applications are desired to run `forever':
a scheduler, a router, a monitor all are not supposed to
terminate but to continue processing the stream of data
presented to them. This has led to the prospering field
of \aname{Data Stream Algorithms}\footnote{which usually
focuses on the (space) complexity of randomized approximations of discrete problems, though}. 
It distinguishes various ways in which the input can be presented to the
program \mycite{Section~4.1}{Muthu}: 
\begin{itemize}
\item[\textbullet] 
In the \emph{Time Series Model}, all data items
(binary digits, say) are to be enumerated in order;
in particular, they must not later be reverted. 
\end{itemize}
This corresponds in TTE to the identity presentation
$\imath$ of an infinite string by itself.
\begin{itemize}
\item[\textbullet]
The \emph{Turnstile Model} on the other hand permits
(finitely many) later updates to previously enumerated 
items. 
\end{itemize}
This corresponds to the presentation
$\imath'$ from Definition~\ref{d:CantorJump}.

\subsection{Revising Output: Terminals} \label{s:Terminal}
Recall the two most basic \texttt{ascii}
control characters understood already by the earliest
text display consoles \cite{Terminal}: \myBS and \myCR.
The first, called ``\emph{backspace}'', moves the cursor
left by one position, thus allowing the last printed
symbol to be overwritten;
whereas the second, ``\emph{carriage return}'', commands
to restart output from the beginning (of the present line).

\begin{example} \label{x:Editing}
The character sequence
\\[0.5ex]\centerline{\tt
G o o d \textvisiblespace \ b y e \myCR 
H e l l o, \textvisiblespace \ M r s \myBS \myBS \myBS w o r l d%
}\\[0.5ex]
will display as: \verb!Hello, world!.
\end{example}
So consider a program generating an infinite sequence of
characters \emph{including} \myBS and \myCR; how do they
appear on an (infinitely long, one-line) display?
Let us require that each character position does settle
down eventually, leading ultimately to the display of a 
truly infinite string (without \myBS and \myCR).

\begin{definition}
A $\myedit$--name of $\bar\sigma\in\Alphabet^\omega$
is an infinite string over $\AlphabetCup{\myCR,\myBS}$
which leads to the display of $\bar\sigma$ in the above sense.
\end{definition}
Now this is exactly what we had already considered in
Definition~\ref{d:CantorJump}:
\begin{observation}
$\myedit\equiv\imath'$.
\end{observation}
Each occurrence of the control character $\myCR$ leads to
the entire display being purged. In order for already the
first character to eventually stabilize, a valid $\myedit$--name 
may thus contain at most finitely many $\myCR$'s.
Let us now consider a terminal incapable of processing $\myBS$,
that is, restrict $\myedit$ to $\AlphabetCup{\myCR}^\omega$.
Then any valid name will make the displayed text settle down
not only character-wise but globally.
This motivates a different
jump operator $\alpha\mapsto\myjump\alpha$
formally introduced in the sequel:

\begin{observation}
$\myedit\big|_{\AlphabetCup{\myCR}^\omega}\equiv\myjump\imath$.
\end{observation}
Hopefully you, most valued reader, are now
indeed curious enough to read on and learn 
about the computational power induced by this

\subsection{Other Kind of Jump} \label{s:HotzJump}
\mycite{Section~5.1}{TOCS} characterizes the computational power of 
\person{Chadzelek} and \person{Hotz}' 
\textsf{quasi-strongly $\delta$--$\IQ$--analytic machines}
in terms of Type-2 machines
by introducing the representation $\myrhoh$ as follows:

\begin{definition} \label{d:RhoHotz}
A $\myrhoh$--name for $x\in\IR$ is a sequence $(q_n)_{_n}$
in $\IQ$ such that
\[  \exists N\in\IN \; \forall n\geq N: \quad |q_n-x|\leq 2^{-n} \enspace \]
\end{definition}
This representation 
is \emph{non-}uniformly equivalent to $\myrho$
yet uniformly (in terms of reducibility that is) 
lies strictly between $\myrho$ and $\myrho'$.

Similarly to Section~\ref{s:Jump}, we now generalize this
particular construction into a generic way:

\begin{definition} \label{d:CantorHotz}
For a representation $\alpha:\subseteq\Alphabet^\omega\to A$,
write $\myjump\alpha:=\alpha\circ\myjump\imath$.

The representation $\myjump\imath:\subseteq\Alphabet^\omega\to\Alphabet^\omega$ in turn
encodes an infinite string $\bar\sigma=(\sigma_n)_{_n}\in\Alphabet^\omega$
as a sequence of infinite strings $\bar\tau_m=(\tau_{(n,m)})_{_n}\in\Alphabet^\omega$, $m\in\IN$,
such that there is some $M\in\IN$ with $\bar\tau_m=\bar\sigma$ for all $m\geq M$.
\end{definition}
In contrast to Definition~\ref{d:CantorJump},
the sequence $(\bar\tau_m)$ is thus required to 
ultimately stabilize \emph{uniformly} in the 
position index $n$.

In view of Claim~f) of the following lemma,
Claims~a) to c) generalize \mycite{Lemma~31}{TOCS};
and Claims~d+e) generalize \cite[\textsc{Proposition~32}b+a]{TOCS}.
\begin{lemma} \label{l:Hotz}
Fix representations $\alpha$ of $A$ and
$\beta$ of $B$.
\begin{enumerate}
\item[a)] An element $a\in A$ is $\alpha$--computable ~if and only if~
  it is $\myjump\alpha$--computable.
\item[b)] It holds ~$\alpha\;\reduceq\;\myjump\alpha\;\reduceq\;\alpha'$.
  The converse reductions are in general discontinuous.
\item[c)] For any function $f:\subseteq A\to B$, 
  $(\alpha\myto\myjump\beta)$--computability
  is equivalent to $(\myjump\alpha\myto\myjump\beta)$--computability.
\item[d)] Every $(\alpha\myto\beta)$--computable function $f$
  is also $(\myjump\alpha\myto\myjump\beta)$--computable; even uniformly in $f$.
\item[e)] An $(\myjump\alpha\myto\myjump\beta)$--computable function
  need not be $(\alpha\myto\beta)$--continuous.
\item[f)] $\myjump\myrho\;\equiv\;\myrhoh$.
\end{enumerate}
\end{lemma}
\begin{proof}
It suffices to treat the case $(A,\alpha)=(B,\beta)=(\Alphabet^\omega,\imath)$---except for f) of course.%
\begin{enumerate}
\item[a)] Encode the $M$ from Definition~\ref{d:CantorHotz}
into the machine computing $(\tau_{(n,m)})_{_{(n,m)}}$ and make
it output $(\tau_{(n,M)})_{_n}$.
\item[b)] The positive claims are immediate,
  the negative ones are straight-forward discontinuity arguments.
\item[c)] 
By a), every $(\imath\myto\myjump\imath)$--computable
function is $(\myjump\imath\myto\myjump\imath)$--computable, too.
For the converse implication, take the Type-2 Machine 
$\calM$
converting $\imath$--names for $x\in\IR$ to
$\myjump\imath$--names for $y=f(x)$. Let $(\bar\sigma_m)$ be
given with $\bar\sigma_m=\bar\sigma_M$ for all $m\geq M$,
$M\in\IN$ unknown.

Now simulate $\calM$ on $\bar\sigma_1$ (implicitly supposing
$M=1$) and simultaneously check that $\bar\sigma_1=\bar\sigma_m$ 
for all $m\geq 1$. 
If (or, rather, when) the latter turns out to fail,
restart under the presumption $M=2$ and so on.
The check will however succeed after finitely many tries
(after reaching the `true' $M$ used in the input).
We thus obtain a finite sequence of output strings,
that is a valid $\myjump\imath$--name for $f(\bar\sigma)$.
\item[d)] 
The \textsf{apply operator} $(F,\bar\sigma)\to F(\bar\sigma)$
is $(\eta^{\omega\omega}\times\myjump\imath\myto\myjump\imath)$--computable:
if $\bar\tau_m=\bar\sigma$ for all $m\geq M$,
then also $F(\bar\tau_m)=F(\bar\sigma)$ for all $m\geq M$.
\item[e)]
Consider the discontinuous
function $F(1^\omega):=1^\omega$, $F(\bar\sigma):=0^\omega$ for $\bar\sigma\not=1^\omega$.
We assert it to be $(\imath\myto\myjump\imath)$--computable;
the claim the follows by c).

Given $\bar\sigma=(\sigma_n)_{_n}$, for each $n=1,2,\ldots$ 
test $\sigma_n=1$ and, as long as this holds, append $1$ to the output.
Otherwise restart the output to $0^\omega$.
Since this restart takes place (if at all) after finite time,
we obtain in either case a valid $\myjump\imath$--name.
\item[f)] Given $(\bar\tau_m)_{_m}$ with $\bar\tau_m=\bar\sigma$ for
  all $m\geq M$, consider for each $m$ the longest initial segment of
  $\tau_m$ constituting the beginning of a valid $\myrho$--name.
  This is computable because $\dom(\myrho)$ is r.e.;
  and it yields a $\myrhoh$--name for $\myrho(\bar\sigma)$,
  i.e. we have ``$\myjump\myrho\reduceq\myrhon$''.
  The converse reduction proceeds similarly.
\qed\end{enumerate}\end{proof}

\section{Real Hypercomputation ~and~ Degrees of Discontinuity} \label{s:Main}
Computability of a function $f:X\to\IR$ in Recursive Analysis means
$(\myrho\myto\myrho)$--computability; equivalently \mycite{Lemma~6.1.7}{Weihrauch}:
the pre-image $f^{-1}[V]=\{x:f(x)\in V\}$
of any open $V\subseteq\IR$ is again open (that is in $\BorelS_1(X)$)
and the pre-image mapping $V\mapsto f^{-1}[V]$ is
$(\thetal\myto\VascoSet{1})$--computable.
In particular, every computable real function is necessarily continuous.

How can we extend the notion of computability to incorporate also
(at least some) discontinuous functions?

\medskip\noindent
Recalling the introduction, one may
\begin{enumerate}
\item[iii)]
  consider $(\myrho\myto\myrhol)$--computable functions.
\end{enumerate}
A function $f:X\to\IR$ is $(\myrho\myto\myrhol)$--continuous 
~if and only if~ it is \emph{lowersemi}continuous, i.e.,
$f^{-1}[V]$ is open for any $V=(y,\infty)$.
It is $(\myrho\myto\myrhol)$--computable
~if and only if~ the mapping
$\IR\ni y\mapsto f^{-1}\big[(y,\infty)\big]\in\BorelS_1(X)$ is 
well-defined and $(\myrhog\myto\VascoSet{1})$--computable 
\cite[\textsc{Theorem~4.5(1)} and \textsc{Corollary~5.1(2)}]{SemiTCS}.
A natural representation (here denoted by $[\myrho\myto\myrhol]$)
of lower-semicontinuous functions on $X$
encodes $f$ as the join of the $\thetal$--names of the open sets
$f^{-1}[(y,\infty)]$, $y\in\IQ$; cf. \mycite{Definition~3.2}{SemiTCS}.

\medskip\noindent
Another approach due to \person{Brattka}, it is equally natural to
\begin{enumerate}
\item[v)]
  consider functions $f:X\to\IR$ for which the
  pre-image $f^{-1}[V]$ of any open $V\subseteq\IR$ belongs to the Borel class
  $\BorelS_d(X)$ (is $\BorelS_d$--measurable) ~ and ~ the mapping $V\mapsto f^{-1}[V]$
  is $(\thetal\myto\VascoSet{d})$--computable
  (called \emph{effectively $\BorelS_d$--measurable}).
\end{enumerate}
The comprehensive paper \cite{EffBorel} thoroughly studies this notion
and its consequences. It is as general as to include also partial
and multi-valued functions on arbitrary computable metric spaces
but in that respect goes beyond our purpose.
\cite{EffBorel} also introduces a natural representation $\VascoFun{d}$ for
$\BorelS_d$--measurable functions as the join of $\VascoSet{d}$--names
of the sets $f^{-1}[V]$, $V$ running through all open rational balls.
For reasons which will be come clear soon, the present work prefers
to write $[\myrho\myto\myrho^{(d-1)}]$ for $\VascoFun{d}$.

\medskip\noindent
Let us unify these two Approaches~iii) and v):
\begin{definition} \label{d:SemiMeasurable}
Call $f:X\to\IR$ be $\BorelS_d$--\textsf{lowersemi}measurable
if $f^{-1}\big[(y,\infty)\big]\in\BorelS_d(X)$ for all $y\in\IR$.
It is \textsf{effectively} $\BorelS_d$--lowersemimeasurable
if $\IR\ni y\mapsto f^{-1}\big[(y,\infty)\big]$ 
is in addition $(\myrhog\myto\VascoSet{d})$--computable.
The representation $[\myrho\myto\myrhol^{(d)}]$ of all
$\BorelS_{d+1}$--lowersemimeasurable functions is defined
to encode $f:X\to\IR$ as the join of $\VascoSet{d}$--names
of $f^{-1}\big[(y,\infty)\big]$ for all $y\in\IQ$.
\end{definition}
Obviously 
$[\myrho\myto\myrho^{(d)}]\equiv[\myrho\myto\myrhol^{(d)}]\myjoin[\myrho\myto\myrhog^{(d)}]$,
exploiting $f^{-1}[U\cap V]=f^{-1}[U]\cap f^{-1}[V]$
and \mycite{Proposition~3.2(4)}{EffBorel} as well as
$\myrho^{(d)}\equiv\myrhol^{(d)}\myjoin\myrhog^{(d)}$
by Lemma~\ref{l:CantorJump}a) and Theorem~\ref{t:JumpRhoD}.

The main result of the present section 
connects these notions to weak function evaluation
$(\myrho\myto\myrho^{(d)})$ and $(\myrho\myto\myrhol^{(d)})$---recall Section~\ref{s:Intro}, Approach~iv)---with 
the representations from Section~\ref{s:JumpRho}.
In fact, justifying the above names for representations of
(lowersemi)measurable functions, we show

\begin{theorem} \label{t:Hypersemi}
\begin{enumerate}
\item[a)]
The \emph{uniformly} characteristic function 
$\cf{}:\BorelS_d(X)\times X\to\{0,1\}$, defined by $(S,\vec x)\mapsto\cf{S}(\vec x):=1$
  if $\vec x\in S$ and $\cf{S}(\vec x):=0$ if $\vec x\not\in S$,
is $(\VascoSet{d}\times\myrho\myto\myrhol^{(d-1)})$--computable.%
\item[b)]
The \textsf{apply operator} $(f,x)\mapsto f(x)$ of $\BorelS_{d+1}$--lowersemimeasurable functions
on $X$ is $\big([\myrho\myto\myrhol^{(d)}]\times\myrho\myto\myrhol^{(d)}\big)$--computable.
\item[c)]
Every $(\myrho\myto\myrhol^{(d)})$--continuous function $f:X\to\IR$
is $\BorelS_{d+1}$--lowersemimeasurable; \\
every $(\myrho\myto\myrhol^{(d)})$--computable one 
is effectively $\BorelS_{d+1}$--lowersemimeasurable, \\
uniformly in $f$ given by an $\eta^{\omega\omega}$--name of a realization.
\end{enumerate}
\end{theorem}
Claims~b) and c) together immediately establish the non-uniform
Theorem~\ref{t:Main}b) which in turn yields Theorem~\ref{t:Main}a).

An alternative proof of Theorem~\ref{t:Main}a), however only for $d\geq3$,
could proceed by induction \mycite{Corollary~9.6}{EffBorel} 
and exploit
that the pointwise limit $f$ of a sequence $f_n$ 
of $(\myrho\myto\myrho^{(d-2)})$--computable functions
is $(\myrho\myto\myrho^{(d-1)})$--computable.

\begin{proof}[Theorem~\ref{t:Hypersemi}]
\begin{enumerate}
\item[a)]
By induction on $d$, starting with $d=1$:
Given a $\myrho$--name of $\vec x\in X$ and a
$\thetal$--name of an open $U\subseteq U$, 
membership ``$\vec x\in U$'' is semi-decidable;
so output 0s while uncertain and start writing 1s
as soon as membership has been established:
this yields a $\myrhol$--name of $\cf{U}(\vec x)$.

Now let $S=\bigcup_n (X\setminus S_n)\in\BorelS_{d+1}(X)$ 
be given by the joint $\VascoSet{(d)}$--names of $S_n\in\BorelS_d(X)$,
$n\in\IN$. By induction hypothesis,
$\myrhol^{(d-1)}$--compute the respective values $y_n:=\cf{S_n}(\vec x)$.
Since $\vec x\in S \Leftrightarrow \exists n:\vec x\not\in S_n$,
we have $\cf{S}(\vec x)=\sup_n (1-y_n)$.
\item[b)]
Given (a $\myrho$--name of) $x\in X$, compute for all $y\in\IQ$ 
a $\VascoSet{d+1}$--name of $S_y:=f^{-1}\big[(y,\infty)\big]$.
Claim~a) yields from that a $\myrhol^{(d)}$--name
of $z_y:=\cf{S_y}(x)$, that is $z_y=1$ in case $x\in S_y$ and
$z_y=0$ in case $x\not\in S_y$. Easy scaling converts that to 
$z_y'=a$ in case $f(x)>y$ and to $z_y'=-\infty$ in case $f(x)\leq y$.
We finally obtain a $\myrhol^{(d)}$--name of $\sup_y z_y'=f(x)$
because $\bar{\IR}^\IN\ni (x_n)_{_n}\mapsto\sup_n x_n\in\bar\IR$ 
is obviously $\big((\myrhol^{(d)})^\IN\myto\myrhol^{(d)}\big)$--computable.
\item[c)]
To start with, recall the proof of \mycite{Theorem~3.7}{SemiTCS} the classical case
\item[$d=0$:]
Evaluate $f$ simultaneously on all $x\in X$ to obtain
rational sequences $p_{x,n}$ with $f(x)=\sup_n p_{x,n}$.
More precisely, using feasible countable (as opposed to
infeasible uncountable) dove-tailing, simulate the machine
evaluating $f$ on all initial parts of $\myrho$--names
of $x\in X$, that is on all finite rational sequences
$\bar q=(q_1,q_2,\ldots,q_N)$ with $N\in\IN$ and
$|q_n-q_k|\leq2^{-n}\forall n\leq k\leq N$.
For each $\bar q$, we obtain as output a 
finite rational sequence 
$(p_{\bar q,m})_{_{m\leq M}}$.
Observe that $\bar q$ is initial segment of a $\myrho$--name
to any $x\in\cball_{\bar q}:=\bigcup_{n=1}^{N(\bar q)}\cball(q_n,2^{-n})$,
$\cball_{\bar q}$ having non-empty interior.
Hence 
\[ 
\exists m: p_{\bar q,m}>a 
\;\quad\Leftrightarrow\quad\;
\forall x\in\cball_{\bar q}:f(x)>a
\;\quad\Leftrightarrow\quad\;
\exists x\in\interior{\cball}_{\bar q}:f(x)>a
\]
which implies
\[
f^{-1}\big[(a,\infty)\big]
\quad=\quad
 \bigcup_{\bar q,m}
  \left\{ \begin{array}{c@{\;:\;}r} \cball_{\bar q} &:\; p_{\bar q,m}>a \\
  \emptyset & p_{\bar q,m}\leq a \end{array}\right\} 
\quad=\quad
 \underbrace{
 \bigcup_{\bar q,m}
  \left\{ \begin{array}{c@{\;:\;}r} \vphantom{\cball_{\bar q}}
  \smash{\interior{\smash{\cball}\vphantom{\ball}}_{\bar q}} &:\; p_{\bar q,m}>a \\
  \emptyset & p_{\bar q,m}\leq a \end{array}\right\}
 }_{\in\BorelS_1}
\]
and immediately yields $\VascoSet{1}$--computability of $f^{-1}[(a,\infty)]$
for given $a\in\IQ$.
\item[${d=1}$:]
Similarly evaluate $f$ on all $x\in X$ to obtain sequences $p_{x,n,m}$
with $\displaystyle f(x)=\mysup_n\myinf_m p_{x,n,m}$. More precisely countable
dove-tailing yields, to each finite $\myrho$--initial segment $\bar q$,
a finite sequence $(p_{\bar q,m,n})_{_{m,n}}$ in $\IQ$ with
\[ 
\exists m\forall n: p_{\bar q,m,n}>a
\;\quad\Leftrightarrow\quad\;
\forall x\in\cball_{\bar q}:f(x)>a
\;\quad\Leftrightarrow\quad\;
\exists x\in\interior{\cball}_{\bar q}:f(x)>a
\]
and hence
\[
f^{-1}\big[(a,\infty)\big]
\quad=\quad
 \underbrace{
 \bigcup_{\bar q,m}
 \overbrace{
 \bigcap_{n}
  \left\{ \begin{array}{c@{\;:\;}r} \cball_{\bar q} &:\; p_{\bar q,m,n}>a \\
  \emptyset & p_{\bar q,m,n}\leq a \end{array}\right\}}^{=:A_{\bar q,m}\in\BorelP_1}
  }_{\in\BorelS_2}
\quad=\quad
 \bigcup_{\bar q,m}
 \bigcap_{n}
  \left\{ \begin{array}{c@{\;:\;}r} \vphantom{\cball_{\bar q}}
  \smash{\interior{\smash{\cball}\vphantom{\ball}}_{\bar q}} &:\; p_{\bar q,m,n}>a \\
  \emptyset & p_{\bar q,m,n}\leq a \end{array}\right\} 
\]
a $\VascoSet{2}$--name of $f^{-1}[(a,\infty)]$ as the $\VascoSet{1}$--names of all
open $X\setminus A_{\bar q,m}$.
\item[${d=2}$:] 
Compute finite rational sequences $(p_{\bar q,m,n,k})_{_{m,n,k}}$ with
\begin{gather*}
\exists m\forall n\exists k: p_{\bar q,m,n,k}>a
\;\quad\Leftrightarrow\quad\;
\forall x\in\cball_{\bar q}:f(x)>a
\;\quad\Leftrightarrow\quad\;
\exists x\in\interior{\cball}_{\bar q}:f(x)>a
\enspace, \\[1ex]
f^{-1}\big[(a,\infty)\big]
\quad=\quad
 \bigcup_{\bar q,m}
 \bigcap_{n}
 \bigcup_k
  \left\{ \begin{array}{c@{\;:\;}r} \cball_{\bar q} & p_{\bar q,m,n,k}>a \\
  \emptyset & p_{\bar q,m,n,k}\leq a \end{array}\right\}
\quad=\quad
 \bigcup_{\bar q,m}
 \underbrace{
 \bigcap_{n}
 \overbrace{
 \bigcup_k
  \left\{ \begin{array}{c@{\;:\;}r} 
  \interior{\smash{\cball}\vphantom{\ball}}_{\bar q} & p_{\bar q,m,n,k}>a \\
  \emptyset & p_{\bar q,m,n,k}\leq a \end{array}\right\}}^{\in\BorelS_1}
  }_{\in\BorelP_2}
\end{gather*}
\item[${d\geq3}$:] analogously.
\qed\end{enumerate}\end{proof}

\section{Power of Type-2 Nondeterminism} \label{s:Nondette}
We now expand on Approach~vi) from the introduction of the present work:
Motivated by \person{B\"{u}chi}'s discovery of \emph{non}deterministic
automata as the appropriate notion of regular languages over \emph{in}finite strings
\cite{Buechi}
as well as by the famous \person{Immerman--Szelepsc\'{e}nyi} concept of nondeterministic
\emph{function} computation \mycite{Theorem~7.6}{Papadimitriou} 
and by \textsf{fair nondeterminism} \cite{Boas}, 
we introduced in \mycite{Section~5}{CIE,TOCS} the \emph{nondeterministic} Type-2 Model:

\begin{definition} \label{d:Nondette}
Let $A$ and $B$ be sets with respective representations
$\alpha:\subseteq\Alphabet^\omega\to A$ and
$\beta:\subseteq\Alphabet^\omega\to B$.
A function $f:\subseteq A\to B$ is called
\emph{nondeterministically} $(\alpha\to\beta)$--computable
if some nondeterministic one-way Turing Machine $\calM$,%
\begin{itemize}
\item upon input of any $\alpha$--name $\bar\sigma\in\Alphabet^\omega$
  for some $a\in\dom(f)$,
\item has a computation which outputs a $\beta$--name
  for $b=f(a)$ \hfill\underline{and}
\item every infinite computation
  of $\calM$ on $\bar\sigma$
  outputs a 
  $\beta$--name for $b=f(a)$.%
\end{itemize}\noindent%
A subset $L$ of $A$ is \emph{nondeterministically decidable}
if the characteristic function 
$\cf{L}:A\to\{0,1\}\times\{\text{\textvisiblespace}\,\}^{\omega}$ 
is nondeterministically $(\alpha\to\imath)$--computable.
\end{definition}
While admittedly even less realistic than a classical $\calNP$--machine,
its capabilities have turned out to exhibit (in addition to
closure under composition) particular structural elegance:
All presentations $\myrho^{(d)}$, $d\in\IN$, can nondeterministically be converted
to and from each other.
Hence we may simply speak of nondeterministic computability and
observe that this notion includes all functions
$(\myrho\myto\myrho^{(d)})$--computable
for \emph{any} $d$, that is by Theorem~\ref{t:Main}a)
the entirety of \person{Brattka}'s hierarchy of effective measurability.

\begin{remark}
In \mycite{Definition~14}{CIE}, we had defined nondeterministic computability
in a way with the third condition in Definition~\ref{d:Nondette}
requiring that any infinite \emph{output} 
of $\calM$ on $\bar\sigma$ constitutes a $\beta$--name for $b=f(a)$.
Since any infinite output requires infinite computation
but not vice versa, this may seem to lead to a different notion.
However both do coincide:
$\calM$ may additionally guess and verify
a function $F:\IN\to\IN$ such that the $n$--th symbol
is output after $F(n)$ steps.
If $F$ has been guessed incorrectly 
(and in particular if, for the given input $\bar\sigma$,
no such $F$ exists at all), then this can be detected
within finite time and abort the computation,
thus complying with the (only seemingly stronger)
Definition~\ref{d:Nondette}.
\end{remark}
The question of exactly characterizing the
power of these machines,
left open in \mycite{Section~5}{TOCS}, is now answered
in terms of the \textsf{Analytical Hierarchy}:

\begin{theorem} \label{t:Nondette}
For $L\subseteq\IN$, the characteristic function 
$\cf{L}:\IN\to\{0,1\}\times\{\text{\textvisiblespace}\,\}^{\omega}$ 
is nondeterministically computable ~if and only if~ $L\in\Delta^1_1$.
\end{theorem}
In particular, the power of Type-2 nondeterminism goes \emph{strictly beyond}
effective measurability; see Corollary~\ref{c:Nondette} below.

The following notion turns out as both natural and useful in the proof
of Theorem~\ref{t:Nondette}:

\begin{definition}
A set $L\subseteq\IN$ is \emph{nondeterministically semi-decidable}
if there exists a nondeterministic Turing machine $\calM$
which, upon input of $x\in\IN$,
\begin{itemize}
\item has a computational path which outputs 
  an infinite string ~ in case $x\in L$;
\item in case $x\not\in L$, aborts after finite time on all computational paths.
\end{itemize}
\noindent
$L$ is \emph{nondeterministically enumerable} if 
a nondeterministic Turing machine $\calM$ without input%
\begin{itemize}
\item has a computational path which outputs
    a list $(x_n)_{_n}$ of integers with $L=\{x_n:n\in\IN\}$;
\item every infinite computation of $\calM$ prints
  a list $(x_n)_{_n}$ of integers with $L=\{x_n:n\in\IN\}$.
\end{itemize}
\end{definition}
Nondeterministic enumerability thus amounts to nondeterministic
computability of an $\operatorname{En}$--name, 
cf. \mycite{Definition~3.1.2.5}{Weihrauch}.
Surprisingly, it turns out as equivalent not to 
nondeterministic \emph{semi-}decidability
but to nondeterministic decidability:

\begin{proposition} \label{p:nondette}
With respect to Type-2 nondeterminism, it holds:
\begin{enumerate}
\item[a)] 
 $\operatorname{En}\equiv\operatorname{Cf}$,~
 where the latter refers to the representation of the powerset of $\IN$
 enumerating a set's members \emph{in order} \mycite{Definition~3.1.2.6}{Weihrauch}.
\item[b)]
 $L\subseteq\IN$ is decidable ~if and only if~ it has a computable $\operatorname{Cf}$--name; \\
 equivalently: both $L$ and its complement are semi-decidable.
\item[c)]
 $L\subseteq\IN$ is semi-decidable ~if and only if~ $L\in\KleeneS_1^1$.
\end{enumerate}
\end{proposition}
\begin{proof}
\begin{enumerate}
\item[a)]
 ``$\operatorname{Cf}\reduceq\operatorname{En}$'' holds already deterministically.
 For the converse we are given a list $(x_n)_{_n}$ of integers enumerating $L$.
 Guess a function $F:\IN\to\IN$ with $x_n\geq m \forall n\geq F(m)$:
 Such obviously $F$ exists; and an incorrect guess can be detected within finite time.
 Knowing $F$, we can determine and sort all restrictions 
 $L\cap[1,m]$, $m\in\IN$. 
\item[b)]
 Immediate.
\item[c)]
 Let $L\in\KleeneS_1^1$. By the 
 \textsf{Normal Form Theorem}---see e.g. \mycite{Proposition~IV.2.5}{Odifreddi}---
 \begin{equation} \label{e:Analytical}
  L \quad=\quad \big\{ x\in\IN \;\big|\; 
    \exists \bar b=(b_n)_{_n}\in\{0,1\}^\omega\;\;
    \forall n\in\IN \;: \; P(x,n,\langle b_1,\ldots,b_n\rangle)\big\}
 \end{equation}
 for some decidable predicate $P$. 
 A nondeterministic Type-2 machine $\calM$, given $x$, may therefore
 guess $\bar b$, check $P(x,n,\bar b|_{\leq n})$ to hold (and output
 a dummy symbol) for each $n\in\IN$ and, when it fails, abort within finite time:
 This yields nondeterministic semi-decision of $L$.

 Conversely let $L$ be semi-decided by $\calM$.
 Then $x\in\IN$ belongs to $L$ ~if and only if~
 there exists a sequence $(b_n)_{_n}$ of guesses $b_n\in\Alphabet^\omega$
 such that $\calM$ makes at last $n$ steps on $x$ and $\bar b$.
 The latter predicate $P(x,n,\langle b_1,\ldots,b_n\rangle)$ being
 decidable, $L$ is of the form (\ref{e:Analytical}).
\qed\end{enumerate}\end{proof}
Claims~b) and c) together yield Theorem~\ref{t:Nondette}.
Moreover we have

\begin{corollary} \label{c:Nondette}
There is nondeterministically computable real $c$
which does not
belong to (any finite\footnote{It may however belong to a transfinite one \cite{Barmpalias}.} 
level of) \person{Weihrauch} and \person{Zheng}'s 
Arithmetical Hierarchy of real numbers.
\end{corollary}
The constant function $f(x)\equiv c$ establishes
Theorem~\ref{t:Main}c).
\begin{proof}[Corollary~\ref{c:Nondette}]
Take some hyperarithmetical but not arithmetical $L\subseteq\IN$,
that is, $L\in\KleeneD_1^1\setminus\KleeneS_0^1$; see e.g.
\mycite{Theorem~\textsection 16.1.XI}{Rogers}
or \mycite{Corollary~IV.2.23}{Odifreddi}.
Since $L$ is nondeterministically decidable, it leads
to a nondeterministically $\myrhob$--computable
real $c:=\sum_{n\in L} 2^{-n}\in\IR$;
compare \mycite{Theorem~4.1.13}{Weihrauch}. 
Were $c$ $\myrho^{(d)}$--computable for some $d\in\IN$,
its (unique!) binary expansion would be decidable
relative to $\emptyset^{(d)}$ \mycite{Theorem~7.8}{Xizhong},
that is in $\KleeneS_{d+1}\subseteq\KleeneS_0^1$, contradiction.
\qed\end{proof}
%

\section{Markov Oracle--Computation} \label{s:Markov}
Returning to Approach~ii) in Section~\ref{s:Intro},
oracle access to the, say, Halting problem does not permit 
computational evaluation $x\mapsto f(x)$ of any discontinuous real function $f$
in the sense of Recursive Analysis, that is with respect
to input $x$ and output $f(x)$ by means of fast convergent
rational sequences.
Other notions of effectivity due to \person{A.A.~Markov, Jr.} 
\mycite{Section~9.6}{Weihrauch} 
and \person{S.~Mazur} \mycite{Section~9.1}{Weihrauch}
restrict real functions to computable arguments $x\in\IRc$.

\begin{definition} \label{d:Markov}
A function $f:\subseteq\IRc\to\IR$ is \emph{Markov--computable} if
is admits a (classically, i.e. discretely) computable Markov realization, 
that is a function $F:\subseteq\IN\to\IN$ such that,
whenever $e$ is G\"{o}del index of a Turing machine $\calM_e$
$\myrho$--computing $x\in\dom(f)$, then $F(e)$ is defined and
index of a machine $\myrho$--computing $f(x)$.
Call $f$  \emph{BM--computable} if
$\big(f(x_n)\big)_{_n}$ is a computable real sequence
whenever $(x_n)_{_n}\in\dom(f)$ is.
\end{definition}
A $(\myrho\myto\myrho)$--computable function is obviously 
Markov--computable which in turn implies BM--computability.
Moreover \person{Mazur}'s theorem asserts every \emph{total}
BM--computable function to be continuous; and
Markov--computability of a \emph{total} real function 
requires $(\myrho\myto\myrho)$--computability 
according to \person{Tseitin} \mycite{Theorem~9.6.6}{Weihrauch}.
See \cite{Peter1,Peter2} for a thorough comparison of all these notions.

\medskip\noindent
Now, as opposed to $(\myrho\myto\myrho)$--computability,
Markov--computability does benefit even topologically from
oracle access:
\begin{example}
The discontinuous sign function $\operatorname{sgn}:\IRc\to\{-1,0,+1\}$ is,
relative to the Halting problem $\emptyset'$,
both Markov--computable and BM--computable.
\end{example}
Observe that in accordance with Definition~\ref{d:Markov}, 
$\operatorname{sgn}$ is considered on the computable reals only.%
\begin{proof}
Given a G\"{o}del index $e$ of some machine $\calM_e$ computing $x$,
modify $\calM_e$ slightly to abort in case $x\not=0$. Feed this new
machine's index $\tilde e$ into Halting oracle. A negative answer
implies $x=0$; the remaining cases $x<0$ and $x>0$ are trivial.
Similarly for BM--computability.
\qed\end{proof}
We are currently working the following generalizations
of Mazur's and Tseitin's Theorems:
\begin{problem}
Fix a total function $f:\IRc\to\IR$.\vspace*{-1ex}%
\begin{enumerate}
\item[a)]
  If $f$ is Markov--computable relative to $\emptyset'$,
  then it is $\BorelS_{2}$--measurable?
\item[b)]
  If $f$ maps every $\myrho'$--computable sequence 
  to a $\myrho'$--computable one, then it is continuous?
\item[c)]
  Characterize the class of total functions
  Markov--computable relative to $\emptyset'$!
\item[d)]
  How about higher degrees?
\end{enumerate}
\end{problem}

\section{Conclusion}
We have characterized $(\myrho\myto\myrho^{(d)})$--computable
functions $f:X\to\IR$
to coincide with \person{Brattka}'s condition of effective
$\BorelS_{d+1}$--measurability;
and shown his representation $\VascoFun{d+1}$ to be natural
for the class of $(\myrho\myto\myrho^{(d)})$--continuous functions.
We furthermore have characterized 
$(\myrho\myto\myrhol^{(d)})$--computable functions and,
extending work of \person{Weihrauch} and \person{Zheng},
found a natural representation for the class of 
$(\myrho\myto\myrhol^{(d)})$--continuous ones.

\begin{problem}
Find a simple characterization of the respective classes of 
$(\myrho^{(k)}\myto\myrho^{(d)})$--continuous,
$(\myrho^{(k)}\myto\myrhol^{(d)})$--continuous,
and $(\myrhol^{(k)}\myto\myrhol^{(d)})$--continuous
functions with $1\leq k\leq d$ arbitrary but fixed;
and devise natural representations for them.
\end{problem}

\medskip\noindent
If $\alpha$ is an \emph{admissible} representation,
then $\alpha^{(d)}$ is usually not for $d\geq1$,
at least not in the strict sense.
This seems to call for \person{Schr\"{o}der}'s 
theory of generalized admissibility \cite{Matthias}.
On the other hand, Corollary~\ref{c:MainTheorems}
succeeded well without this notion.

\paragraph{Acknowledgments:}
The author is grateful to \person{K.~Weihrauch}, 
\person{V.~Brattka}, \person{P.~Hertling}, and \person{X.~Zheng}
for constant support and seminal discussions.
Section~\ref{s:Nondette} results from research initiated
by \person{U.~Kohlenbach} and \person{D.~Norman} during 
\textsf{CiE'05}. I would also like to repeat the many thanks
expressed in Footnote\ref{f:Thanks}.


\end{document}